\centerline{\bf Distributed Order Reaction-Diffusion Systems Associated with Caputo Derivatives}

\vskip.3cm\centerline{R.K. Saxena$^{a}$, A.M. Mathai$^{b,c}$, and H.J. Haubold$^{b,d}$}

\vskip.2cm\noindent $^{a}$ Department of Mathematics and Statistics, Jai Narain Vyas University, Jodhpur-342004, India
\vskip.2cm\noindent $^{b}$ Centre for Mathematical Sciences, Pala, Kerala-686574, India
\vskip.2cm\noindent $^{c}$ Department of Mathematics and Statistics, McGill University, Montreal, Canada
\vskip.3cm\noindent $^{d}$ Office for Outer Space Affairs, United Nations, Vienna International Centre, 1400-Vienna, Austria
\vskip.5cm\noindent
{\bf Abstract}
\vskip.3cm
\noindent
This paper deals with the investigation of the solution of an unified fractional reaction-diffusion equation of distributed order associated with the Caputo derivatives as the time-derivative and Riesz-Feller fractional derivative as the space-derivative. The solution is derived by the application of the joint Laplace and Fourier transforms in compact and closed form in terms of the H-function. The results derived are of general nature and include the results investigated earlier by other authors, notably by Mainardi et al. [23,24], for the fundamental solution of the space-time fractional equation, including Haubold et al. [13] and Saxena et al. [38] for fractional reaction-diffusion equations. The advantage of using the Riesz-Feller derivative lies in the fact that the solution of the fractional reaction-diffusion equation, containing this derivative, includes the fundamental solution for space-time fractional diffusion, which itself is a generalization of fractional diffusion, space-time fraction diffusion, and time-fractional diffusion. These specialized types of diffusion can be interpreted as spatial probability density functions evolving in time and are expressible in terms of the H-function in compact forms. The convergence conditions for the double series occurring in the solutions are investigated. It is interesting to observe that the double series comes out to be a special case of the Srivastava-Daoust hypergeometric function of two variables given in the Appendix B of this paper.


\vskip.5cm\noindent{\bf 1.\hskip.3cm Introduction}

\vskip.3cm\noindent In recent papers, several authors have demonstrated the applications of reaction-diffusion models in pattern formation in biology, chemistry, and physics, in this connection, refer to Metzler and Klafter [26], Murray [28], Kuramoto [21], Wilhelmsson and Lazzaro [43], Hundsdorfer and Verwer [19], Gafiyuchuk [9,10], Guo and Xu [12], Chen et al. [2], Engler [5], Pagnini et al. [33], and Huang and Liu [18]. These systems indicate that diffusion can produce the spontaneous formation of spatio-temporal patterns. For details, see the work of Cross and Hohenberg [3], and Nicolis and Prigogine [30]. A general model for reaction-diffusion systems is investigated by Henry and Wearne [15,16], Henry et al. [17], Diethelm [4], Saxena et al. [37,38,39,40,41], Mathai et al.[25], and Mainardi [22].

\vskip.2cm The object of this paper is to derive the solution of an unified model of reaction-diffusion system (2.1), associated with the Caputo derivative as the time-derivative and the Riesz-Feller derivative as the space-derivative. This new model provides the extension of the models discussed earlier by authors, including the models discussed by Pagnini and Mainardi [33]. Fractional order sub-diffusion is discussed by Naber [29].

\vskip.5cm\noindent{\bf 2.\hskip.3cm Solution of Unified Fractional Reaction-Diffusion Equations}

\vskip.3cm\noindent {\bf Theorem 1.}\hskip.3cm{\it Consider the one-dimensional unified fractional reaction-diffusion equation of distributed order
$$\eqalignno{{_0D}_t^{\alpha}N(x,t)+a~{_0D}_t^{\beta} N(x,t)&=\lambda~{_xD}_{\theta}^{\gamma} N(x,t)+U(x,t)&(2.1)\cr
x\in R,~t>0,~0<\alpha\le 1,~0<\beta\le 1,& 0<\gamma\le 2&(2.2)\cr
\noalign{\hbox{with initial conditions}}
N(x,0)&=f(x),~x\in R, \lim_{x\rightarrow\pm\infty}N(x,t)=0,~t>0,&(2.3)\cr}
$$where $\lambda$ is a diffusion constant, ${_0D_t}^{\alpha}$ and ${_0D_t}^{\beta}$ are the Caputo derivative operators of orders $\alpha$ and $\beta$ respectively,  ${_xD_{\theta}}^{\gamma}$ is the Riesz-Feller derivative of order $\gamma$ and skewness $\theta$; $|\theta|\le \min\{\gamma,2-\gamma\}$ and $U(x,t)$ is a nonlinear function which belongs to the area of reaction-diffusion, then there holds the following formula for the solution of (2.1):

$$\eqalignno{N(x,t)&=\sum_{r=0}^{\infty}{{(-a)^r}\over{2\pi}}\int_{-\infty}^{\infty}f^{*}(k)\exp(-ikx)\cr
&\times\left[E_{\alpha,(\alpha-\beta)r+1}^{r+1}(-bt^{\alpha})+a~t^{\alpha-\beta}
E_{\alpha,(\alpha-\beta)(r+1)+1}^{r+1}(-bt^{\alpha})\right]{\rm d}k\cr
&+\sum_{r=0}^{\infty}{{(-a)^r}\over{2\pi}}\int_0^t\xi^{\alpha+(\alpha-\beta)r-1}
\int_{-\infty}^{\infty}U^{*}(k,t-\xi)\exp(-ikx)\cr
&\times E_{\alpha,\alpha+(\alpha-\beta)r}^{r+1}(-b\xi^{\alpha}){\rm d}k~{\rm d}\xi,&(2.4)\cr}
$$where $\Re (\alpha)>0,~\Re (\beta)>0,~\Re(\alpha-\beta)>0$ and $b=\lambda~{\Psi}_{\gamma}^{\theta}(k)$.}

\vskip.3cm\noindent{\bf Proof}. \hskip.3cm If we apply the Laplace transform with respect to the time variable $t$, Fourier transform with respect to space variable $x$ and use the initial conditions (2.2), (2.3) and the formula (A3) and (A14), then the given equation transforms into the form

$$s^{\alpha}{\tilde{ N}}^{*}(k,s)-s^{\alpha-1}f^{*}(k)+as^{\beta}{\tilde{ N}}^{*}(k,s)-as^{\beta-1}f^{*}(k)=-\lambda~\Psi_{\gamma}^{\theta}(k){\tilde{ N}}^{*}(k,s)+{\tilde{ U}}^{*}(k,s),
$$where, according to the convention followed, the symbol $\tilde{(\cdot)}$ will stand for the Laplace transform with respect to the time variable $t$ and ${*}$ represents the Fourier transform with respect to the space variable $x$. Solving for ${\tilde N}^{*}(k,s)$ yields
$${\tilde N}^{*}(k,s)={{f^{*}(k)[s^{\alpha-1}+as^{\beta-1}]}\over{s^{\alpha}+as^{\beta}+\lambda~\Psi_{\gamma}^{\theta}(k)}}+{{{\tilde U}^{*}(k)}\over{s^{\alpha}+ax^{\beta}+\lambda~\Psi_{\gamma}^{\theta}(k)}},\eqno(2.5)
$$where $b=\lambda~\psi_{\gamma}^{\theta}(k)$. On taking the inverse Laplace transform of (2.5) and applying the formula (A14), it is found that

$$\eqalignno{N^{*}(k,t)&=f^{*}(k)\bigg[\sum_{r=0}^{\infty}(-a)^{r}t^{(\alpha-\beta)r}
E_{\alpha,(\alpha-\beta)r+1}^{r+1}(-bt^{\alpha})\cr
&+at^{\alpha-\beta}\sum_{r=0}^{\infty}(-a)^{r}t^{(\alpha-\beta)r}
E_{\alpha,(\alpha-\beta)(r+1)+1}^{r+1}(-bt^{\alpha})\bigg]\cr
&+\int_0^tU^{*}(k,t-\xi)\sum_{r=0}^{\infty}(-a)^r\xi^{\alpha+(\alpha-\beta)r-1}
E_{\alpha,\alpha+(\alpha-\beta)r}^{r+1}(-b\xi^{\alpha}){\rm d}\xi.&(2.6)\cr}
$$The required solution (2.4) is now obtained by taking the inverse Fourier transform of (2.6). This completes the proof of Theorem 1.

\vskip.5cm\noindent{\bf Alternative form of the solution (2.4)}

\vskip.3cm\noindent By using the series representation of the generalized Mittag-Leffler function $E_{\beta,\gamma}^{\alpha}(z)$ in (A15), the expression

$$\eqalignno{t^{\alpha-\rho}\sum_{r=0}^{\infty}(-a)^rt^{(\alpha-\beta)r}&E_{\alpha,\alpha
+(\alpha-\beta)r-\rho+1}^{r+1}(-bt^{\alpha}),\cr
\noalign{\hbox{can be written as}}
t^{\alpha-\rho}\sum_{r=0}^{\infty}\sum_{u=0}^{\infty}{{(1)_{r+u}}\over{r!u!}}
&{{(-at^{\alpha-\beta})^r(-bt^{\alpha})^u}\over{\Gamma(\alpha-\rho+1+(\alpha-\beta)r
+\alpha u)}}\cr
&=t^{\alpha-\rho}S_{1:0;0}^{1:0;0}\left[-at^{\alpha-\beta},-bt^{\alpha}\bigg\vert_{
[\alpha-\rho+1:\alpha-\beta;\alpha]:-;-}^{[1:1;1]:-;-}\right],&(2.7)\cr}
$$where $S(\cdot)$ is the Srivastava-Daoust hypergeometric function of two variables [42]. The definition of this function is given in Appendix B. Hence, Theorem 1 can be stated in terms of the Srivastava-Daoust hypergeometric function of two variables in the following form: Under the conditions of Theorem 1, the one-dimensional fractional reaction-diffusion equation

$$\eqalignno{{_0D}_t^{\alpha}N(x,t)+a~{_0D}_t^{\beta}N(x,t)
&=\lambda~{_xD}_{\theta}^{\gamma}N(x,t)+U(x,t),&(2.8)\cr
\noalign{\hbox{has the solution given by}}
N(x,t)&={{1}\over{2\pi}}\int_0^{\infty }f^{*}(k)\exp(-ikx)\bigg[S_{1:0;0}^{1:0;0}\left[-at^{\alpha-\beta},
-bt^{\alpha}\bigg\vert_{[1:\alpha-\beta;\alpha]:-;-}^{[1:1;1]:-;-}\right]\cr
&+at^{\alpha-\beta}S_{1:0;0}^{1:0;0}\left[-at^{\alpha-\beta},-bt^{\alpha}
\bigg\vert_{[\alpha:\alpha-\beta;\alpha]:-;-}^{[1:1;1]:-;-}\right]\bigg]{\rm d}k\cr
&+{{1}\over{2\pi}}\int_0^t\xi^{\alpha-1}\int_{-\infty}^{\infty}U^{*}(k,t-\xi)\exp(-ikx)\cr
&\times S_{1:0;0}^{1:0;0}\left[-at^{\alpha-\beta},-bt^{\alpha}\bigg\vert_{[\alpha:\alpha-\beta;
\alpha]:-;-}^{[1:1;1]:-;-}\right]{\rm d}k~{\rm d}\xi,&(2.9)\cr}
$$where $\Re(\alpha)>0,~\Re(\beta)>0,~\Re(\alpha-\beta)>0$ and
$b=\lambda~{_x\Psi}_{\gamma}^{\theta}(k)$.

\vskip.3cm\noindent Note 1.\hskip.3cm By virtue of the Lemma given in Appendix B, the double infinite power series occurring in Theorem 1 converge for $\Re(\alpha)>0,~\Re(\alpha-\beta)>0$.

\vskip.3cm\noindent{\bf 3.\hskip.3cm Special Cases}

\vskip.3cm\noindent When $f(x)=\delta(x)$, where $\delta(x)$ is the Dirac-delta function, we obtain the following

\vskip.3cm\noindent{\bf Corollary 1.1.}\hskip.3cm{\it Consider the one-dimensional fractional reaction-diffusion equation of distributed order
$${_0D}_t^{\alpha}N(x,t)+a~{_0D}_t^{\beta}N(x,t)=\lambda~{_xD}_{\theta}^{\gamma}N(x,t)
+U(x,t)\eqno(3.1)
$$for $x\in R,~0<\alpha\le 1,~0<\beta\le 1, ~t>0$ with initial conditions
$$N(x,0)=\delta(x),~x\in R,~\lim_{x\rightarrow\pm\infty}N(x,t)=0,~t>0,\eqno(3.2)
$$where $\lambda$ is a diffusion constant, $U(x,t)$ is a nonlinear function which belongs to the area of reaction-diffusion, then there holds the following formula for the fundamental solution of (3.1):

$$\eqalignno{N(x,t)&=\sum_{r=0}^{\infty}{{(-a)^r}\over{2\pi}}\int_{-\infty}^{\infty}
t^{(\alpha-\beta)r}\exp(-ikx)\cr
&\times [E_{\alpha,(\alpha-\beta)r+1}^{r+1}(-bt^{\alpha})+at^{\alpha-\beta}
E_{\alpha,\alpha+(\alpha-\beta)(r+1)+1}^{r+1}(-bt^{\alpha})]{\rm d}k\cr
&+\sum_{r=0}^{\infty}{{(-a)^r}\over{2\pi}}\int_0^t\xi^{\alpha+(\alpha-\beta)r-1}
\int_{-\infty}^{\infty}U^{*}(k,t-\xi)\exp(-ikx)E_{\alpha,\alpha+
(\alpha-\beta)r}^{r+1}(-b\xi^{\alpha}){\rm d}k~{\rm d}\xi,&(3.3)\cr}
$$where $\Re(\alpha)>0,~\Re(\beta)>0,~\Re(\alpha-\beta)>0$ and $b=\lambda~{\Psi_{\gamma}}^{\theta}(k)$.}
\vskip.2cm
If we set $\theta=0$ in (3.1), the Riesz-Feller derivative reduces to Riesz fractional derivative defined by (A8) and it yields the following result given by Saxena et al. [39]:

\vskip.3cm\noindent{\bf Corollary 1.2.}\hskip.3cm{\it Consider the one-dimensional unified fractional reaction-diffusion equation

$$\eqalignno{{_0D}_t^{\alpha}N(x,t)+a~{_0D}_t^{\beta}N(x,t)&=\lambda~{_xD}_0^{\gamma}N(x,t)
+U(x,t)&(3.4)\cr
\noalign{\hbox{for $x\in R,~t>0,~0<\alpha\le 1,~0<\beta\le 1$ with initial conditions}}
N(x,0)&=f(x),~x\in R, ~\lim_{x\rightarrow\pm\infty}N(x,t)=0,~t>0,&(3.5)\cr}
$$where $\lambda$ is a diffusion constant, ${_0D}_t^{\alpha}$ and ${_0D}_t^{\beta}$ are the Caputo derivative operators of orders $\alpha$ and $\beta$ respectively, ${_xD}_0^{\gamma}$ is the Riesz fractional derivative, $U(x,t)$ is nonlinear function which belongs to the area of reaction-diffusion, then there holds the following formula for the solution of (3.4).

$$\eqalignno{N(x,t)&=\sum_{r=0}^{\infty}{{(-a)^r}\over{2\pi}}\int_{-\infty}^{\infty}
t^{(\alpha-\beta)r}f^{*}(k)\exp(-ikx)\cr
&\times [E_{\alpha,(\alpha-\beta)r+1}^{r+1}(-\pi|k|^{\gamma}t^{\alpha})
+a~t^{\alpha-\beta}E_{\alpha,(\alpha-\beta)(r+1)+1}^{r+1}(-\lambda|k|^{\gamma}t^{\alpha})]{\rm d}k\cr
&+\sum_{r=0}^{\infty}{{(-a)^r}\over{2\pi}}\int_0^t\xi^{\alpha+(\alpha-\beta)r-1}
\int_{-\infty}^{\infty}U^{*}(k,t-\xi)\exp(-ikx)E_{\alpha,\alpha+(\alpha-\beta)r}^{r+1}
(-\lambda|k|^{\gamma}\xi^{\alpha}){\rm d}k~{\rm d}\xi,&(3.6)\cr}
$$where $\Re(\alpha)>0,~\Re(\beta)>0,~\Re(\alpha-\beta)>0.$}

\vskip.2cm If we further take $f(x)=\delta(x)$, the above corollary reduces to the following result:

\vskip.3cm\noindent{\bf Corollary 1.3.}\hskip.3cm{\it Consider the one-dimensional unified fractional reaction-diffusion equation
$$\eqalignno{{_0D}_t^{\alpha}N(x,t)+a~{_0D}_t^{\beta}N(x,t)
&=\lambda~{_xD}_{\theta}^{\gamma}N(x,t)+U(x,t)&(3.7)\cr
\noalign{\hbox{for $x\in R, 0<\alpha\le 1,~0<\beta\le 1,~t>0$, with initial conditions}}
N(x,0)&=\delta(x),~x\in R, ~\lim_{x\rightarrow\pm\infty}N(x,t)=0,~t>0,&(3.8)\cr}
$$where $\lambda$ is a diffusion constant, $U(x,t)$ is nonlinear function which belongs to the area of reaction-diffusion, then there holds the following formula for the fundamental solution of (3.7).

$$\eqalignno{N(x,t)&=\sum_{r=0}^{\infty}{{(-a)^r}\over{2\pi}}
\int_{-\infty}^{\infty}t^{(\alpha-\beta)r}\exp(-ikx)\cr
&\times [E_{\alpha,(\alpha-\beta)r+1}^{r+1}(-\lambda|k|^{\gamma}t^{\alpha})+a~t^{\alpha-\beta}
E_{\alpha,(\alpha-\beta)(r+1)+1}^{r+1}(-\lambda|k|^{\gamma}t^{\alpha})]{\rm d}k\cr
&+\sum_{r=0}^{\infty}{{(-a)^r}\over{2\pi}}\xi^{\alpha+(\alpha-\beta)r-1}
\int_{-\infty}^{\infty}U^{*}\exp(-ikx)E_{\alpha,\alpha+(\alpha-\beta)r+1}^{r+1}
(-\lambda|k|^{\gamma}\xi^{\alpha}){\rm d}k~{\rm d}\xi,&(3.9)\cr}
$$where $\alpha>\beta$.}

\vskip.3cm If we set $a=0$, then the theorem reduces to the following result given by Haubold et al. [13, p.684].

\vskip.3cm\noindent{\bf Corollary 1.4.}\hskip.3cm{\it Consider the one-dimensional unified fractional reaction-diffusion equation
$$\eqalignno{{_0D}_t^{\alpha}N(x,t)&=\lambda~{_xD}_{\theta}^{\gamma}N(x,t)+U(x,t)&(3.10)\cr
\noalign{\hbox{for $x\in R,~t>0,~0<\alpha\le 1,~0<\beta\le 1,~0<\gamma\le 2$ with initial conditions}}
N(x,0)&=f(x),~x\in R,~\lim_{x\rightarrow\pm\infty}N(x,t)=0,~t>0,&(3.11)\cr}
$$where $\lambda$ is a diffusion constant, and $U(x,t)$ is a nonlinear function which belongs to the area of reaction-diffusion, then there holds the following formula for the solution of (3.10).

$$\eqalignno{N(x,t)&={{1}\over{2\pi}}\int_{-\infty}^{\infty}f^{*}(k)
E_{\alpha,1}(-\lambda t^{\alpha}\Psi_{\gamma}^{\theta}(k))\exp(-ikx){\rm d}k\cr
&+{{1}\over{2\pi}}\int_0^t\xi^{\alpha-1}\int_{-\infty}^{\infty}U^{*}(k,t-\xi)
E_{\alpha,\alpha}(-\lambda\Psi_{\gamma}^{\theta}(k))\exp(-ikx){\rm d}k~{\rm d}\xi,&(3.12)\cr}
$$where $E_{\alpha,\alpha}(z)$ is the Mittag-Leffler function.}

\vskip.3cm Now if we set $f(x)=\delta(x), ~\gamma=2,~\theta=0,~\alpha $ replaced by $2\alpha$ and $\beta$ by $\alpha$, and make use of (A18) then the following result is obtained:

\vskip.3cm\noindent{\bf Corollary 1.5.}\hskip.3cm{\it Consider the following one-dimensional reaction-diffusion system
$$\eqalignno{{{\partial^{2\alpha}N(x,t)}\over{\partial t^{2\alpha}}}+a{{\partial^{\alpha}N(x,t)}\over{\partial t^{\alpha}}}&=\nu^2{{\partial^2N(x,t)}\over{\partial x^2}}+U(x,t),~0<\alpha\le 1&(3.13)\cr
\noalign{\hbox{with the initial conditions}}
N(x,0)&=\delta(x),~x\in R,~\lim_{x\rightarrow\pm\infty}N(x,t)=0,~t>0,&(3.14)\cr}$$ where $U(x,t)$ is a nonlinear function belonging to the area of reaction-diffusion. Then for the fundamental solution of (3.13), subject to the initial condition (3.14), there holds the formula

$$\eqalignno{N(x,t)&={{1}\over{2\pi\sqrt{(a^2-4b)}}}\int_{-\infty}^{\infty}\exp(-ikx)
[(\sigma+a)E_{\alpha}(\sigma~t^{\alpha})-(\mu+a)E_{\alpha}(\mu t^{\alpha})]{\rm d}k\cr
&+{{1}\over{2\pi}}\int_0^t\xi^{\alpha-1}\int_{-\infty}^{\infty}\exp(-ikx)U^{*}(k,t-\xi)
[E_{\alpha,\alpha}(\sigma\xi^{\alpha})-E_{\alpha,\alpha}(\mu\xi^{\alpha})]{\rm d}k~{\rm d}\xi,&(3.15)\cr}
$$where $\sigma$ and $\mu$ are the real and distinct roots of the quadratic equation $y^2+ay+b=0$, given by
$$\sigma={1\over2}(-a+\sqrt{(a^2-4b)})\hbox{ and }\mu={1\over2}(-a-\sqrt{(a^2-4b)}),\eqno(3.16)
$$where $b^2=\nu^2k^2$.}

\vskip.3cm Next, if we further set $U(x,t)=0$, we then obtain the following result which includes many known results on fractional telegraph equations, including the one recently given by Orsingher et al. [32]:

\vskip.3cm\noindent{\bf Corollary 1.6.}\hskip.3cm{\it Consider the following one-dimensional reaction-diffusion system

$$\eqalignno{{{\partial^{2\alpha}N(x,t)}\over{\partial t^{2\alpha}}}+a{{\partial^{\alpha}N(x,t)}\over{\partial t^{\alpha}}}&=\nu^2{{\partial^2N(x,t)}\over{\partial x^2}},~0\le\alpha\le 1&(3.17)\cr
\noalign{\hbox{with the initial conditions}}
N(x,0)&=\delta(x),~\in R,~\lim_{x\rightarrow\pm\infty}N(x,t)=0,~t>0.&(3.18)\cr}
$$Then for the fundamental solution of (3.17), subject to the initial conditions (3.18), there holds the formula
$$N(x,t)={{1}\over{2\pi\sqrt{(a^2-4b)}}}\int_{-\infty}^{\infty}\exp(-ikx)[(\sigma+a)
E_{\alpha}(\sigma t^{\alpha})-(\mu+a)E_{\alpha}(\mu t^{\alpha})]{\rm d}k,\eqno(3.19)
$$where $\sigma$ and $\mu$ are defined in (3.16), $b=\nu^2k^2$, and $E_{\alpha}(x)$ is the Mittag-Leffler function defined in (A17).}

\vskip.3cm The result (3.19) can be rewritten in the explicit form as

$$\eqalignno{N(x,t)&={{1}\over{4\pi}}\int_{-\infty}^{\infty}\exp(-ikx)
[(1+{{a}\over{\sqrt{(a^2-4\nu^2k^2)}}})E_{\alpha}(\sigma t^{\alpha})\cr
&+(1-{{a}\over{\sqrt{(a^2-4\nu^2 k^2)}}})E_{\alpha}(\mu t^{\alpha})]{\rm d}k,&(3.20)\cr}
$$where $\sigma$ and $\mu$ are defined in (3.16) and $E_{\alpha}(x)$ is the Mittag-Leffler function defined in (A17).

\vskip.2cm The equation (3.20) represents the solution of the time-fractional telegraph equation (3.17) subject to the initial conditions (3.18), recently solved by Orsingher et al. [32]. It is remarked here that the solution given by Orsingher et al. [32] is in terms of the Fourier transform of the solution in the form given below. It is observed that the Fourier transform of the solution of the equation (3.17) can be expressed in the form

$$N^{*}(x,t)={1\over2}\left\{(1+{{a}\over{\sqrt{(a^2-4\nu^2k^2)}}})E_{\alpha}(\sigma t^{\alpha})+(1-{{a}\over{\sqrt{(a^2-4\nu^2k^2)}}})E_{\alpha}(\mu t^{\alpha})\right\}\eqno(3.21)
$$

\vskip.3cm\noindent{\bf 4.\hskip.3cm An Additional Type of Reaction-Diffusion Equation}

\vskip.3cm\noindent{\bf Theorem 2.}\hskip.3cm{\it Under the conditions of Theorem 1 with $0<\alpha\le 1$ replaced by $1<\alpha<2$, $N(x,0)=f(x)$ and $N_t(x)=g(x),$ and following a similar procedure the solution of the following reaction-diffusion equation

$$\eqalignno{{_0D}_t^{\alpha}N(x,t)&+a~{_0D}_t^{\beta}N(x,t)=\lambda~{_xD}_{\theta}^{\gamma}N(x,t)+U(x,t)&(4.1)\cr
\noalign{\hbox{is givn by}}
N(x,t)&=\sum_{r=0}^{\infty}{{(-a)^r}\over{2\pi}}\int_{-\infty}^{\infty}t^{(\alpha-\beta)r}
f^{*}(k)\exp(-ikx)\cr
&\times [E_{\alpha,(\alpha-\beta)r+1}^{r+1}(-t^{\alpha})+at^{\alpha-\beta}
E_{\alpha,(\alpha-\beta)(r+1)+1}^{r+1}(-t^{\alpha})]{\rm d}k\cr
&+\sum_{r=0}^{\infty}{{(-a)^r}\over{2\pi}}\int_{-\infty}^{\infty}t^{(\alpha-\beta)r}g^{*}(k)\exp(-ikx)\cr
&\times[tE_{\alpha,(\alpha-\beta)r+2}^{r+1}(-t^{\alpha-1})+at^{\alpha-\beta+1}E_{\alpha,(\alpha-\beta)(r+1)+2}^{r+1}(-t^{\alpha})\cr
&+\sum_{k=0}^{\infty}{{(-a)^r}\over{2\pi}}\int_0^t\xi^{\alpha+(\alpha-\beta)r-1}
\int_{-\infty}^{\infty}U^{*}(k,t-\xi)\exp(-ikx)\cr
&\times E_{\alpha,\alpha+(\alpha-\beta)r}^{r+1}(-b\xi^{\alpha})]{\rm d}k~{\rm d}\xi,&(4.2)\cr}
$$where $\Re(\alpha)>0,~\Re(\beta)>0,~\Re(\alpha-\beta)>0$ and $b=\lambda~{_x\Psi}_{\gamma}^{\theta}(k)$.}

\vskip.3cm\noindent{\bf Corollary 2.1.}\hskip.3cm{\it Under the conditions of Theorem 2 with $\theta=0$, the solution of the following reaction-diffusion equation

$$\eqalignno{{_0D}_t^{\alpha}&+a~{_0D}_t^{\beta}N(x,t)
=\lambda~{_xD}_{\theta}^{\gamma}N(x,t)+U(x,t)&(4.3)\cr
\noalign{\hbox{is given by}}
N(x,t)&=\sum_{r=0}^{\infty}{{(-a)^r}\over{2\pi}}\int_{-\infty}^{\infty}t^{(\alpha-\beta)r}
f^{*}(k)\exp(-ikx)\cr
&\times[E_{\alpha,(\alpha-\beta)r+1}^{r+1}(-\lambda|k|^{\gamma}t^{\alpha})
+at^{\alpha-\beta}E_{\alpha,(\alpha-\beta)(r+1)+1}^{r+1}(-\lambda|k|^{\gamma}t^{\alpha})]
{\rm d}k\cr
&+\sum_{r=0}^{\infty}{{(-a)^r}\over{2\pi}}\int_{-\infty}^{\infty}t^{(\alpha-\beta)r}
g^{*}(k)\exp(-ikx)\cr
&\times[tE_{\alpha,(\alpha-\beta)r+1}^{r+1}(-\lambda|k|^{\gamma}t^{\alpha-1})
+t^{\alpha-\beta+1}E_{\alpha,(\alpha-\beta)(r+1)+2}^{r+1}(-\lambda|k|^{\gamma}t^{\alpha})]\cr
&+\sum_{r=0}^{\infty}{{(-a)^r}\over{2\pi}}\int_0^t\xi^{\alpha+(\alpha-\beta)r-1}
\int_{-\infty}^{\infty}U^{*}(k,t-\xi)\exp(-ikx)\cr
&\times E_{\alpha,\alpha+(\alpha-\beta)r}^{r+1}(-\lambda|k|^{\gamma}\xi^{\alpha}){\rm d}k~{\rm d}\xi,&(4.4)\cr}
$$where $\Re(\alpha)>0,~\Re(\beta)>0,~\Re(\alpha-\beta)>0$.}

\vskip.3cm\noindent{\bf Corollary 2.2.}\hskip.3cm{\it Under the conditions of Theorem 2 with $0<\alpha\le 1$ replaced by $1<\alpha<2$ and $f(x)=g(x)=\delta(x)$, where $\delta(x)$ is the Dirac-delta function, the fundamental solution of the following reaction-diffusion equation

$$\eqalignno{{_0D}_t^{\alpha}N(x,t)&+a~{_0D}_t^{\beta}N(x,t)=\lambda~{_xD}_{\theta}^{\gamma}N(x,t)+U(x,t)&(4.5)\cr
\noalign{\hbox{is given by}}
N(x,t)&=\sum_{r=0}^{\infty}{{(-a)^r}\over{2\pi}}\int_{-\infty}^{\infty} t^{(\alpha-\beta)r}\exp(-ikx)\cr
&\times [E_{\alpha,(\alpha-\beta)r+1}^{r+1}(-bt^{\alpha})+a~t^{\alpha-\beta}
E_{\alpha,(\alpha-\beta)(r+1)+1}^{r+1}(-b t^{\alpha})]{d}k\cr
&+\sum_{r=0}^{\infty}{{(-a)^r}\over{2\pi}}\int_{-\infty}^{\infty}t^{(\alpha-\beta)r}
\exp(-ikx)\cr
&\times[tE_{\alpha,(\alpha-\beta)r+2}^{r+1}(-bt^{\alpha-1})+at^{\alpha-\beta+1}
E_{\alpha,(\alpha-\beta)(r+1)+2}^{r+1}(-bt^{\alpha})]\cr
&+\sum_{r=0}^{\infty}{{(-a)^r}\over{2\pi}}\int_0^t\xi^{\alpha+(\alpha-\beta)r-1}
\int_{-\infty}^{\infty}U^{*}(k,t-\xi)\exp(-ikx)\cr
&\times E_{\alpha,\alpha+(\alpha-\beta)r}^{r+1}(-b\xi^{\alpha}){\rm d}k~{\rm d}\xi,&(4.6)\cr}
$$where $\Re(\alpha)>0,~\Re(\beta)>0,~\Re(\alpha-\beta)>0,$ and $b=\lambda~{\Psi_{\gamma}}^{\theta}(k)$.

\vskip.2cm As a concluding remark, it is observed that Theorem 2 also holds further if instead of one Riesz-Feller derivative, we consider a finite number of Feller derivatives. This result is given in Theorem 3 in the next section.

\vskip.3cm\noindent{\bf 5.\hskip.3cm Several Feller Derivatives}

\vskip.3cm\noindent{\bf Theorem 3.}\hskip.3cm {\it Consider the following one-dimensional unified reaction-diffusion equation of fractional order
$${_0D}_t^{\alpha}N(x,t)+a~{_0D}_t^{\beta}N(x,t)=[\sum_{j=1}^{m}
\lambda_j~{_xD}_{\theta_j}^{\gamma_j}N(x,t)]+U(x,t),~m\in N\eqno(5.1)
$$for $x\in R,~t>0,~0<\alpha\le 2,~0<\beta\le 2,~0<\gamma_j\le 2, j=1,...,m$ with initial conditions
$$N(x,0)=f(x),~N_t(x,0)=g(x),~x\in R,~\lim_{x\rightarrow\pm\infty}N(x,t)=0,~t>0,\eqno (5.2)
$$where $\lambda_j>0,~j=1,...,m$ are diffusion constants, ${_0D}_t^{\alpha}$ and ${_0D}_t^{\beta}$ are the Caputo derivative operators of orders $\alpha$ and $\beta$ respectively, ${_xD}_{\theta_j}^{\gamma_j}, j=1,...,m$ are the Riesz-Feller fractional derivatives of orders $\gamma_j$ and skewness $|\theta_j|\le \min_{1\le j\le m}[\gamma_j,2-\gamma_j],~j=1,...,m$, respectively. Then there holds the following formula for the solution of (5.1):
$$\eqalignno{N(x,t)&=\sum_{r=0}^{\infty}{{(-a)^r}\over{2\pi}}\int_{-\infty}^{\infty}
t^{(\alpha-\beta)r}f^{*}(k)\exp(-ikx)\cr
&\times [E_{\alpha,(\alpha-\beta)r+1}^{r+1}(-b^{*}t^{\alpha})+t^{\alpha-\beta+1}
E_{\alpha,(\alpha-\beta)(r+1)+1}^{r+1}(-b^{*}t^{\alpha})]{\rm d}k\cr
&+\sum_{r=0}^{\infty}{{(-a)^r}\over{2\pi}}\int_{-\infty}^{\infty}t^{(\alpha-\beta)r}
g^{*}(k)\exp(-ikx)\cr
&\times[tE_{\alpha,(\alpha-\beta)r+2}^{r+1}(-b^{*}t^{\alpha})+a~t^{\alpha-\beta+1}
E_{\alpha,(\alpha-\beta)(r+1)+2}^{r+1}(-b^{*}t^{\alpha})]\cr
&+\sum_{r=0}^{\infty}{{(-a)^r}\over{2\pi}}\int_0^t\xi^{\alpha+(\alpha-\beta)r-1}
\int_{-\infty}^{\infty}U^{*}(k,t-\xi)\exp(-ikx)\cr
&\times E_{\alpha,\alpha+(\alpha-\beta)r}^{r+1}(-b^{*}\xi^{\alpha}){\rm d}k~{\rm d}\xi,&(5.3)\cr}
$$where $\Re(\alpha)>0,~\Re(\beta)>0,~\Re(\alpha-\beta)>0$ and $b^{*}=\sum_{j=1}^m\lambda_j~\Psi_{\gamma_j}^{\theta_j}(k)$.}

\vskip.5cm\noindent{\bf References}

\vskip.3cm\noindent[1] Caputo, M.: {\it Elasticita e Dissipazione},  Zanichelli, Bologna 1969.
\vskip.2cm\noindent[2] Chen, J., Liu, R., Turner, I., and Anh, V.: The fundamental and numerical solutions of the Riesz space-fractional reaction-dispersion equation, {\it The Australian and New Zealand Industrial and Applied Mathematics Journal (ANZIAM)}, {\bf 50} (2008), 45-57.

\vskip.2cm\noindent[3] Cross, M.C. and Hohenberg, P.C.: Pattern formation outside of equilibrium, {\it Reviews of Modern Physics}, {\bf 65} (1993), 851-912.

\vskip.2cm\noindent[4] Diethelm, K.: {\it The Analysis of Fractional Differential Equations}, Springer, Berlin 2010.

\vskip.2cm\noindent[5] Engler, H.: On the speed of spread for fractional reaction-diffusion, {\it International Journal of Differential Equations}, 2010, Article ID 315421, 16 pages.

\vskip.2cm\noindent[6] Erd\'elyi, A., Magnus, W., Oberhettinger, F., and Tricomi, F.G.: {\it Higher Transcendental Functions}, {\bf Vol. 3}, McGraw-Hill, New York 1954.

\vskip.2cm\noindent[7] Feller, W.: On a generalization of Marcel Riesz potentials and the semi-groups generated by them, {\it Meddelanden Lunds Universitets Matematiska Semiarium (Comm. S\'em. Math\'em. Uniersit\'e de Lund)}, {\bf Tome Suppl. D\'edi\'e M. Riesz, Lund} (1952), 73-81.

\vskip.2cm\noindent[8] Feller, W.: {\it An Introduction to Probability Theory and Its Applications}, {\bf Vol.2}, Second Edition, Wiley, New York 1971 (First Edition 1966).

\vskip.2cm\noindent[9] Gafiychuk, V., Datsko, B., and Meleshko, V.: Mathematical modeling of pattern formation in sub- and superdiffusive reaction-diffusion systems, 2006, arXiv:nlin/0611005.

\vskip.2cm\noindent[10] Gafiychuk, V., Datsko, B. and Meleshko, V.: Nonlinear oscillations and stability domains in fractional reaction-diffusion systems, 2007, arXiv:nlin/0702013.

\vskip.2cm\noindent[11] Gorenflo, R. and Mainardi, F.: Approximation of L\'evy-Feller diffusion by random walk, {\it Journal for Analysis and Its Applications}, {\bf 18} (1999), 1-16.

\vskip.2cm\noindent[12] Guo, X. and Xu, M.: Some physical applications of Schr\"odinger equation, {\it Journal of Mathematical Physics}, 47082104 (2008): doi10, 1063/1.2235026, 9 pages.

\vskip.2cm\noindent[13] Haubold, H.J., Mathai, A.M., and Saxena, R.K.: Solutions of reaction-diffusion equations in terms of the H-function, {\it Bulletin of the Astronomical Society, India}, {\bf 35}, (2007), 681-689.

\vskip.2cm\noindent[14] Haubold, H.J., Mathai, A.M., and Saxena, R.K.: Further solutions of reaction-diffusion equations in terms of the H-function, {\it Journal of Computational and Applied Mathematics}, {\bf 235} (2011), 1311-1316.

\vskip.2cm\noindent[15] Henry, B.I. and Wearne, S.: Fractional reaction-diffusion, {\it Physica A},{\bf 276} (2000), 448-455.

\vskip.2cm\noindent[16] Henry, B.I. and Wearne, S.: Existence of Turing instabilities in a two species reaction-diffusion system, {\it SIAM Journal of Applied Mathematics}, {\bf 62} (2002), 870-887.

\vskip.2cm\noindent[17] Henry, B.I., Langlands, T.A.M., and Wearne, S.L.: Turing pattern formation in fractional activator-inhibitor systems, {\it Physical Review E} {\bf 72} (2005), 026101, 14 pages.

\vskip.2cm\noindent[18] Huang, F. and Liu, R.: The time-fractional diffusion equation and the advection dispersion equation, {\it The Australian and New Zealand Industrial and Applied Mathematics Journal (ANZIAM)}, {\bf 46} (2005), 1-14.

\vskip.2cm\noindent[19] Hundsdorfer, W. and Werver, J.C.: {\it Numerical Solution of Time-dependent Advection-Diffusion-Reaction Equations}, Springer, Berlin 2003.

\vskip.2cm\noindent[20] Kilbas, A.A., Srivastava, H.M., and Trujillo, J.J.: {\it Theory and Applications of Fractional Differential Equations}, Elsevier, Amsterdam 2006.

\vskip.2cm\noindent[21] Kuramoto, Y.: {\it Chemical Oscillation, Waves and Turbulence}, Dover Publications, Mineola, New York 2003.

\vskip.2cm\noindent[22] Mainardi, F.: {\it Fractional Calculus and Waves in Linear Viscoelasticity}, Imperial College Press, London 2010.

\vskip.2cm\noindent[23] Mainardi, F. Luchko, Y., and Pagnini, G.: The fundamental solution of the space-time fractional diffusion equation, {\it Fractional Calculus and Applied Analysis}, {\bf 4} (2001), 153-202.

\vskip.2cm\noindent[24] Mainardi, F.,  Pagnini, G., and Saxena, R.K.: Fox H-functions in fractional diffusion, {\it Journal of Computational and  Applied Mathematics}, {\bf 178} (2005), 321-331.

\vskip.2cm\noindent[25] Mathai, A.M., Saxena, R.K., and Haubold, H.J.: {\it The H-function: Theory and Applications}, Springer, New York 2010.

\vskip.2cm\noindent[26] Metzler, R. and Klafter, J.: The random walk's guide to anomalous diffusion: A fractional dynamics approach, {\it Physics Reports}, {\bf 339} (2000), 1-77.

\vskip.2cm\noindent[27] Miller, K.S. and Ross, B.: {\it An Introduction to the Fractional Calculus and Fractional Differential Equations}, Wiley, New York 1993.

\vskip.2cm\noindent[28] Murray, J.D.: {\it Mathematical Biology}, Springer, New York 2003.

\vskip.2cm\noindent[29] Naber, M.: {\it Distributed order fractional sub-diffusion}, {\it Fractals}, {\bf 12} (2004), 23-32.

\vskip.2cm\noindent[30] Nicolis, G. and Prigogine, I.: {\it Self-Organization in Nonequilibrium Systems: From Dissipative Structures to Order through Fluctuations}, Wiley, New York 1977.

\vskip.2cm\noindent[31] Nikolova, Y. and Boyadjiev, L.: Integral transform methods to solve a time-space fractional diffusion equation, {\it Fractional Calculus and Applied Analysis}, {\bf 13} (2010), 57-67.

\vskip.2cm\noindent[32] Orsingher, F. and Beghin, L.: Time-fractional telegraph equations and telegraph processes with Brownian time, {\it Probability Theory and Related Fields}, {\bf 128} (2004), 141-160.

\vskip.2cm\noindent[33] Pagnini, R. and Mainardi, F.: Evolution equations for the probabilistic generalization of Voigt profile function, {\it Journal of Computational and Applied Mathematics}{\bf 233} (2010),1590-1595.

\vskip.2cm\noindent[34] Podlubny, I. {\it Fractional Differential Equations}, Academic Press, New York, 1959.

\vskip.2cm\noindent[35] Prabhakar, T.R.: A singular integral equation with a generalized Mittag-Leffler function in the kernel, {\it Yokohama Mathematical Journal}, {\bf 19} (1971), 7-15.

\vskip.2cm\noindent[36] Samko, S.G., Kilbas, A.A., and Marichev, O.I.: {\it Fractional Integrals and Derivatives: Theory and Applications}, Gordon and Breach Science Publishing, Switzerland 1993.

\vskip.2cm\noindent[37] Saxena, R.K., Saxena, R., and Kalla, S.L.: Computational solution of a fractional generalization of Schr\"odinger equation occurring in quantum mechanics, {\it Applied Mathematics and Computation}, {\bf 216}  (2010), 1412-1417.

\vskip.2cm\noindent[38] Saxena, R.K., Mathai, A.M., and Haubold, H.J.: Fractional reaction-diffusion equations, {\it Astrophysics and Space Science}, {\bf 305} (2006a), 289-296.

\vskip.2cm\noindent[39] Saxena, R.K., Mathai, A.M., and Haubold, H.J.: Reaction-diffusion systems and nonlinear waves, {\it Astrophysics and Space Science}, {\bf 305} (2006b), 297-303.

\vskip.2cm\noindent[40] Saxena, R.K., Mathai, A.M., and Haubold, H.J.: Solution of generalized fractional reaction-diffusion equations, {\it Astrophysics and Space Science}, {\bf 305} (2006c), 305-313.

\vskip.2cm\noindent[41] Saxena, R.K., Mathai, A.M., and Haubold, H.J.: Solution of fractional reaction-diffusion equations in terms of the Mittag-Leffler functions, {\it International Journal of Scientific Research}, {\bf 15} (2006d), 1-17.

\vskip.2cm\noindent[42] Srivastava, H.M. and Daoust, M.C.: A note on the convergence of Kamp\'e de F\'eriet double hypergeometric series, {\it Mathematische Nachrichten}, {\bf 53} (1972), 151-159.

\vskip.2cm\noindent[43] Wilhelmsson, H. and Lazzaro, E.: {\it Reaction-Diffusion Problems in the Physics of Hot Plasmas}, Institute of Physics Publishing, Bristol and Philadelphia 2001.

\vskip.2cm\noindent[44] Wiman, A.: Ueber den Fundamentalsatz in der Theorie der Funktionen $E_{\alpha}(x)$, {\it Acta Mathematica}, {\bf 29} (1955a), 191-201.

\vskip.2cm\noindent[45] Wiman, A.: Ueber die Nullstellen der Funktionen $E_{\alpha}(x)$ {\it Acta Mathematica}, {\bf 29} (1955b), 217-268.

\vskip.5cm\noindent{\bf Appendices}

\vskip.3cm\noindent{\bf Appendix A: Caputo and Riesz-Feller fractional derivatives}
\vskip.3cm The following fractional derivative of order $\alpha>0$ is introduced by Caputo [1] in the form

$$\eqalignno{{_0D}_t^{\alpha}f(x,t)&={{1}\over{\Gamma(m-\alpha)}}\int_0^t{{f^{(m)}(x,\tau)}\over{(t-\tau)^{\alpha+1-m}}}{\rm d}\tau,~m-1<\alpha<m,~\Re(\alpha)>0,~m\in N&(A1)\cr
&={{\partial^m f(x,t)}\over{\partial t^m}},\hbox{  if  }\alpha=m,&(A2)\cr}
$$where ${{\partial^m}\over{\partial t^m}}f(x,t)$ is the $m$-th partial derivative of $f(x,t)$ with respect to $t$. The Laplace transform of the Caputo derivative is given by Caputo [1] (also see Podlubny [34], Kilbas et al. [20]) in the form
$$L\{{_0D}_t^{\alpha}f(x,t);s\}=s^{\alpha}F(x,s)-\sum_{r=0}^{m-1}s^{\alpha-r-1}f^{(r)}(x,0_{+}),~m-1<\alpha\le m,\eqno(A3)
$$where $F(x,s)$ is the Laplace transform of $f(x,t)$ with respect to $t$. This derivative is useful in the solutions of applied problems connected with anomalous reaction, anomalous diffusion, and anomalous reaction-diffusion problems, which are expressible in terms of partial fractional differential equations. In this connection, one can refer to the monograph by Podlubny [34], Samko et al. [36], Miller and Ross [27], Kilbas et al. [20], Mainardi [22], Diethelm [4], and recent papers on the subject by Nikolova [31], Naber [29], and Pagnini et al. [33].
\vskip.2cm
Following Feller [7,8] it is conventional to define the Riesz-Feller space-fractional derivative of order $\alpha$ and skewness $\theta$ in terms of its Fourier transform in the form:
$$\eqalignno{F\{{_xD}_{\theta}^{\alpha}f(x);k\}&=-\psi_{\alpha}^{\theta}(k)f^{*}(k),&(A4)\cr
\noalign{\hbox{where $f^{*}(k)$ denotes the Fourier transform of $f(x)$ with respect to $x$}}
\psi_{\alpha}^{\theta}(k)&=|k|^{\alpha}\exp(i(sign~k){{\theta\pi}\over{2}}),~0<\alpha\le 2,~|\theta|\le\min\{\alpha,2-\alpha\}.&(A5)\cr}
$$Further, when $\theta=0$, we have a symmetric operator with respect to $x$ that can be interpreted as

$${_xD}_0^{\alpha}=-\left[-{{{\rm d}^2}\over{{\rm d}x^2}}\right]^{\alpha/2}.\eqno(A6)
$$This can be formally deduced by writing $-(k)^{\alpha}=-(k^2)^{\alpha/2}$. For $0<\alpha<2$ and $|\theta|\le\min\{\alpha,2-\alpha\}$, the Riesz-Feller derivative can be shown to possess the following integral representation in the  $x$ domain:

$$\eqalignno{{_xD}_{\theta}^{\alpha}f(x)&={{\Gamma(1+\alpha)}\over{\pi}}\{\sin[(\alpha+\theta)\pi/2]
\int_0^{\infty}{{f(x+\xi)-f(x)}\over{\xi^{1+\alpha}}}{\rm d}\xi\cr
&+\sin[(\alpha-\theta)\pi/2]\int_0^{\infty}{{f(x-\xi)-f(x)}\over{\xi^{1+\alpha}}}{\rm d}\xi\}.&(A7)\cr}
$$For $\theta=0$, the Riesz-Feller fractional derivative becomes the Riesz fractional derivative of order $\alpha$ for $1<\alpha\le 2$ defined by analytic continuation in the whole range $0<\alpha\le 2,~\alpha\ne 1$, see Gorenflo and Mainardi [11], as

$$\eqalignno{{_xD}_0^{\alpha}&=-\lambda[I_{+}^{-\alpha}-I_{-}^{-\alpha}],&(A8)\cr
\noalign{\hbox{where}}
\lambda&={{1}\over{2\cos(\alpha\pi/2)}};~~I_{\pm}^{-\alpha}={{{\rm d}^2}\over{{\rm d}x^2}}I_{\pm}^{2-\alpha}.&(A9)\cr}
$$The Weyl fractional integral operators are defined in the monograph by Samko et al. [36] as
$$\eqalignno{(I_{+}^{\beta}N)(x)&={{1}\over{\Gamma(\beta)}}\int_{-\infty}^x(x-\zeta)^{\beta-1}N(\zeta){\rm d}\zeta,~\beta>0\cr
(I_{-}^{\beta}N)(x)&={{1}\over{\Gamma(\beta)}}\int_x^{\infty}(\zeta-x)^{\beta-1}N(\zeta){\rm d}\zeta,~\beta>0.&(A10)\cr}$$

\noindent Note 1.  We note that ${_xD}_0^{\alpha}$ is a pseudo differential operator. In particular we have
$${_xD}_0^2={{{\rm d}^2}\over{{\rm d}x^2}},~\hbox{  but  }{xD}_0^{1}\ne {{{\rm d}}\over{{\rm d}x}}.\eqno(A11)
$$For $\theta=0$ we have
$$F\{{_xD}_0^{\alpha}f(x);k\}=-|k|^{\alpha}f^{*}(k).\eqno(A12)
$$We also need the following result in the analysis that follows: Haubold et al. [13] have shown that
$$F^{-1}[E_{\beta,\gamma}(-at^{\beta}\Psi_{\alpha}^{\theta}(k);x]={{1}\over{\alpha|x|}}
H_{3,3}^{2,1}\left[{{|x|}\over{(at^{\beta})^{1/\alpha}}}\bigg\vert_{(1,{{1}\over{\alpha}},
(1,1),(1,\rho)}^{(1,{{1}\over{\alpha}}),(\gamma,{{\beta}\over{\alpha}}),(1,\rho)}\right],\eqno(A13)
$$where $\rho={{\alpha-\theta}\over{2\alpha}}$,  $\Re(\alpha)>0,~\Re(\beta)>0,~\Re(\gamma)>0$. The following results given by Saxena et al. [39] are also required in the analysis that follows: From [39] we have
$$L^{-1}\left\{{{s^{\rho-1}}\over{s^{\alpha}+as^{\beta}+b}};t\right\}=t^{\alpha-\rho}
\sum_{r=0}^{\infty}(-a)^rt^{\alpha-\beta)r}E_{\alpha,\alpha+(\alpha-\beta)r-\rho+1}^{r+1}(-bt^{\alpha}),\eqno(A14)
$$where $\Re(\alpha)>0,~\Re(\beta)>0,~\Re(\alpha-\beta)>0,~\Re(\alpha-\rho)>0,~\Re(s)>0,~|{{as^{\beta}}\over{s^{\alpha}+b}}|<1$ and $E_{\beta,\gamma}^{\alpha}(z)$ is the generalized Mittag-Leffler function of Prabhakar [35], defined by
$$E_{\beta,\gamma}^{\alpha}(z)=\sum_{n=0}^{\infty}{{(\alpha)_nz^n}\over{\Gamma(n\beta+\gamma)n!}}\eqno(A15)
$$for $\alpha,\beta,\gamma\in C, \Re(\beta)>0,~\Re(\gamma)>0$, where the Pochhammer symbol $(\alpha)_n$ is defined by
$$(\alpha)_n=\alpha(\alpha+1)...(\alpha+n-1),~(\alpha)_0=1,~\alpha \ne 0,~(\alpha)_n={{\Gamma(\alpha+n)}\over{\Gamma(\alpha)}}
$$whenver $\Gamma(\alpha)$ is defined. When $\alpha=1$ in (A15), the expression reduces to the generalized Mittag-Leffler function, defined by Wiman [44,45] as
$$E_{\beta,\gamma}(z)=\sum_{n=0}^{\infty}{{z^n}\over{\Gamma(n\beta+\gamma)}},~\Re(\beta)>0,~\Re(\gamma)>0.\eqno(A16)
$$When $\gamma=1$, (A16) reduces to the Mittag-Leffler function (see, Erd\'elyi et al.[6])
$$E_{\beta}(z)=\sum_{n=0}^{\infty}{{z^n}\over{\Gamma(n\beta+1)}},~\Re(\beta)>0.\eqno(A17)
$$We also have (see, Saxena et al.[39])

$$L^{-1}\left[{{s^{2\alpha-1}+as^{\alpha-1}}\over{s^{2\alpha}+as^{\alpha}+b}}\right]
={{1}\over{\sqrt{(a^2-4b)}}}[(\sigma+a)E_{\alpha}(\sigma t^{\alpha})-(\mu+a)E_{\alpha}(\mu t^{\alpha})],\eqno(A18)
$$where $\Re(\alpha)>0,~\Re(s)>0$ and $\sigma$ and $\mu$ are the real and distinct roots of the quadratic equation $x^2+ax+b=0$, namely,

$$\sigma={1\over2}(-a+\sqrt{(a^2-4b)})\hbox{ and }\mu={1\over2}(-a-\sqrt{(a^2-4b)}).\eqno(A19)
$$

\vskip.3cm\noindent{\bf Appendix B: Convergence of the double power series}

\vskip.3cm\noindent{\bf Lemma}.\hskip.3cm{\it For all $a,\alpha,\beta>0$, there holds the formula

$$\sum_{m,n\ge 0}{{(1)_{m+n}}\over{m!n!}}{{x^my^n}\over{(a)_{\alpha m+\beta n}}}=\Gamma(a)S_{1:0;0}^{1:1;1}\left(x,y\bigg\vert_{[a:\alpha;\beta];-;-}^{[1:1;1];-;-}\right),\eqno(B1)
$$where $S$ stands for the Srivastava-Daoust function (Srivastava et al. [42]).}

\vskip.3cm The Srivastava-Daoust generalization of the Kamp\'e de F\'eriet hypergeometric series in two variables is defined by the double hypergeometric series as [42, p.151];

$$\eqalignno{S_{C:D;D'}^{A:B;B'}(x,y)&=S_{C:D;D'}^{A:B;B'}\left[x,y\bigg\vert_{[(c):
\delta;\epsilon]:[(d):\eta]:[(d'):\eta']}^{[(a):\theta,\Phi]:[(b):\Psi]:[(b'):\Psi']}\right]\cr
&=\sum_{m=0}^{\infty}\sum_{n=0}^{\infty}g_{m,n}{{x^my^n}\over{m!n!}}\cr
\noalign{\hbox{where}}
g_{m,n}&={{\{\prod_{j=1}^A\Gamma(a_j+m\theta_j+n\Phi_j)\}\{\prod_{j=1}^B\Gamma(b_j+m\Psi_j)\}
\{\prod_{j=1}^{B'}\Gamma(b_j'+n\Psi_j')\}}\over{\{\prod_{j=1}^C\Gamma(c_j+m\delta_j+n\epsilon_j)\}
\{\prod_{j=1}^D\Gamma(d_j+m\eta_j)\}\{\prod_{j=1}^{D'}\Gamma(d_j'+n\eta_j')\}}}&(B2)\cr}
$$with the coefficients $\theta_1,...,\theta_A,...,\eta_1',...,{\eta_{D'}}' >0$. For the sake of brevity $(a)$ is taken to denote the sequence of $A$ parameters $a_1,...,a_A$ with similar interpretations for $(b),...,(d')$. Srivastava and Daoust have shown, [42, p.155], that the series (B2) converges for all $x,y\in C$ when

$$\eqalignno{\Delta&=1+\sum_{j=1}^C\delta_j+\sum_{j=1}^D\eta_j-\sum_{j=1}^A\theta_j-\sum_{j=1}^B\Psi_j>0,&(B3)\cr
\Delta'&=1+\sum_{j=1}^C\epsilon_j+\sum_{j=1}^{D'}{\eta_j}'-\sum_{j=1}^A\Phi_j-\sum_{j=1}^{B'}{\Psi_j}'>0.&(B4)\cr}
$$For a detailed account of the convergence conditions of the double series, the reader is referred to the original paper by Srivastava and Daoust [42].
\vskip.2cm Now the left side of (B1) can be written as

$$\sum_{m=0}^{\infty}\sum_{n=0}^{\infty}{{(1)_{m+n}}\over{m!n!}}{{x^my^n}\over{(a)_{\alpha m+\beta n}}}=\Gamma(a)\sum_{m=0}^{\infty}\sum_{n=0}^{\infty}{{\Gamma(1+m+n)}\over{\Gamma(a+\alpha m+\beta n)}}{{x^my^n}\over{m!n!}}\eqno(B5)
$$which establishes (B1), since for the expression in (B1), $\Delta=\alpha,~\Delta'=\beta$.

\bye